\documentclass[11pt]{article} 

\usepackage[utf8]{inputenc} 

\usepackage{cite}

\usepackage{hyperref}
\hypersetup{colorlinks=false}

\usepackage{geometry,amsmath} 
\usepackage{graphicx} 

\numberwithin{equation}{section}

\usepackage{booktabs} 
\usepackage{array} 
\usepackage{paralist} 
\usepackage{verbatim} 
\usepackage{subfig} 

\usepackage{float}

\usepackage{todonotes}

\usepackage{amssymb}
\usepackage{amsmath}
\usepackage{amsthm}
\usepackage{mathrsfs}

\usepackage{xcolor,pict2e}

\theoremstyle{plain}
\newtheorem{thm}{Theorem}[section]

\newtheorem{rem}[thm]{Remark}

\usepackage{fancyhdr} 
\usepackage{sectsty}
\allsectionsfont{\sffamily\mdseries\upshape} 

\usepackage[nottoc,notlof,notlot]{tocbibind} 
\usepackage[titles,subfigure]{tocloft} 

\title{Symmetries and reductions of integrable nonlocal partial differential equations}
\author{Linyu Peng\vspace{0.4cm}
\\
{\it Waseda Institute for Advanced Study, Waseda University, Tokyo 169-8050, Japan}\\
{\it Email: l.peng@aoni.waseda.jp}}

\date{} 

\begin{document}
\maketitle

\abstract{In this paper, symmetry analysis is extended to study nonlocal differential equations, in particular  two integrable nonlocal equations,  the nonlocal nonlinear Schr\"odinger equation and the nonlocal modified Korteweg--de Vries equation.  Lie point symmetries are obtained based on a general theory and used to reduce these equations to nonlocal and local ordinary differential equations separately; namely one symmetry may allow reductions to both nonlocal and local equations depending on how the invariant variables are chosen. For the nonlocal modified Korteweg--de Vries equation, analogously to the local situation,  all reduced local equations are integrable. 
At the end, we also define complex transformations to connect nonlocal differential  equations and differential-difference equations.
\vspace{0.2cm}


}

\section{Introduction}
Symmetries have been fundamentally important for understanding solutions of  differential equations, e.g. \cite{Ol1993,Hy2000,BlKu1989,AcHe1975,BlCo1969}. It also reveals the integrability of partial differential equations; for instance,  the Ablowitz--Ramani--Segur conjecture states that every ordinary differential equation obtained by an exact reduction of an integrable evolution equation solvable via inverse scattering transforms is of the P-type, namely without movable critical points, c.f. \cite{AbRaSe1978}. In this paper, the powerful symmetry techniques are extended to study nonlocal differential equations with space and/or time reflections. It can not only provide insights for obtaining analytic solutions, but also reveal integrability of  nonlocal equations. After writing down  a general theory in Section \ref{sec:theory}, two integrable nonlocal differential equations, the nonlocal nonlinear Schr\"odinger (NLS) equation \cite{AbMu2013} and the nonlocal modified Korteweg--de Vries  (mKdV) equation \cite{JiZh2017}, are separately investigated as illustrative examples. The results are immediately applicable to the many nonlocal differential equations proposed during the recent years, e.g. \cite{Ya2015,Fo2016,AbMu2017,LoHu2017,SoXiZh2017a}.
 
The nonlocal NLS equation 
\begin{equation}\label{eq:noSc0}
\operatorname{i}q_t(x,t)+q_{xx}(x,t)+2q^2(x,t)q^{\ast}(-x,t)=0,
\end{equation}
 was derived by Ablowitz and Musslimani \cite{AbMu2013} via a reduction of the AKNS system. The nonlocal NLS equation admits a great number of good properties that the classical NLS equation possesses, e.g. PT-symmetric, admitting Lax-pair and infinitely many conservation laws, solvable using inverse scattering transforms. During the latest years, integrable nonlocal systems have received great attention with many newly-proposed models, e.g. nonlocal vector NLS equation \cite{Ya2015}, multidimensional extension of nonlocal NLS equation \cite{Fo2016}, nonlocal sine-Gordon equation, nonlinear derivative NLS equation and many other systems \cite{AbMu2017}, nonlocal mKdV equation \cite{JiZh2017}, Alice--Bob physics \cite{LoHu2017}, nonlocal Sasa--Satsuma equation \cite{SoXiZh2017a}, to mention only a few. Solutions of these systems have also been explored by many scholars; see for example,  \cite{AbMu2016,GuPe2018,JiZh2017,KhSa2015,SoXiZh2017a,SoXiZh2017b,Zh2018,XuCh2016}.

One contrast as Ablowitz and Musslimani noticed, e.g.\cite{AbMu2013,AbMu2017}, is that reductions of nonlocal equations amount to nonlocal ordinary differential equations (ODEs), e.g. nonlocal Painlev\'e-type equations. In this paper, we show that alternative ways allow us to avoid such inconvenience. We first classify all Lie point symmetries of the nonlocal NLS equation \eqref{eq:noSc0} and the nonlocal mKdV equation \cite{JiZh2017}
\begin{equation}
u_t(x,t)+u(x,t)u(-x,-t)u_x(x,t)+u_{xxx}(x,t)=0.
\end{equation}
Possible symmetry reductions are then conducted for both equations. We find out that one may reduce an nonlocal differential equation to both nonlocal and local ODEs by choosing the invariant variables in different ways. In other words, we are able to kill all nonlocal terms in the reduced ODEs by choosing the invariant variables in a proper manner. In particular for the nonlocal mKdV equation, all reduced local ODEs are integrable. These results are included in Section \ref{sec:NLS} and Section \ref{sec:mKdV}.  In Section \ref{sec:ddes}, simple transformations are defined to connect nonlocal differential equations with differential-difference equations (DDEs).

\section{The linearized symmetry condition for nonlocal differential equations }
\label{sec:theory}
We first introduce the multi-index notations needed for the symmetry techniques of local differential equations, e.g. \cite{Ol1993}, that will be extended to nonlocal differential equations. 

Let $x=(x^1,x^2,\ldots, x^m)\in \mathbb{R}^m$ be the independent variables and let $u=(u^1,u^2,\ldots,u^n)\in \mathbb{R}^n$  be the dependent variables. Note that in many occasions, people also tend to use $(x,t)$ to denote independent variables as the space $x$ and time $t$; this convention will occur in the next sections but for now we are happy without distinguishing one another.  Partial derivatives of $u^{\alpha}$ are written in the multi-index form $u_{\bold{J}}^{\alpha}$ where $\bold{J}=(j_1,j_2,\ldots,j_m)$ with each index $j_i$ a non-negative integer denoting  the number of derivatives with respect to $x^i$. Namely
\begin{equation}
u^{\alpha}_{\bold{J}}=\frac{\partial^{|\bold{J}|} u^{\alpha}}{\partial (x^1)^{j_1}\partial (x^2)^{j_2}\ldots \partial (x^m)^{j_m}},
\end{equation}
where $|J|=j_1+j_2+\cdots +j_m$.  Consider a one-parameter group of Lie point transformations as follows:
\begin{equation}\label{eq:lpt}
\widetilde{x}=\widetilde{x}(\varepsilon;x,u),\quad \widetilde{u}=\widetilde{u}(\varepsilon;x,u),
\end{equation}
subject to $\widetilde{x}|_{\varepsilon=e}=x$, $\widetilde{u}|_{\varepsilon=e}=u$. Here $\varepsilon=e$ is the identity element of the one-parameter group.
Define the total derivative with respect to $x^i$ as 
\begin{equation}
D_i=\frac{\partial}{\partial x^i}+\sum_{\alpha,J}u_{\bold{J}+\bold{1}_i}^{\alpha}\frac{\partial}{\partial u_{\bold{J}}^{\alpha}},
\end{equation}
where $\bold{1}_i$ is the
$m$-tuple with only one nonzero entry $1$ in the $i$-th place. 
The notation $\partial_{x^i}=\frac{\partial}{\partial x^i}$ and so forth will also be used.
The corresponding infinitesimal generator is 
\begin{equation}\label{eq:inge}
\bold{v}=\xi^i(x,u)\partial_{x^i}+\phi^{\alpha}(x,u)\partial_{u^{\alpha}},
\end{equation}
where 
\begin{equation}
\xi^i=\frac{\operatorname{d}\! \widetilde{x}^i}{\operatorname{d}\!\varepsilon}\Big|_{\varepsilon=e},\quad \phi^{\alpha}=\frac{\operatorname{d}\!\widetilde{u}^{\alpha}}{\operatorname{d}\! \varepsilon}\Big|_{\varepsilon=e}.
\end{equation}
Note that the Einstein summation convention is used here and elsewhere if necessary. The prolonged generator $\operatorname{pr}\bold{v}$ can be written in terms of $u$, $\xi$, $\phi$ and their derivatives:
\begin{equation}
\operatorname{pr}\bold{v}=\bold{v}+\sum_{\alpha,|\bold{J}|\geq1}\phi_{\bold{J}}^{\alpha}(x,[u])\partial_{u_{\bold{J}}^{\alpha}},
\end{equation}
where $[u]$ is shorthand for $u$ and finitely many of its partial derivatives and the coefficients are recursively given by 
\begin{equation}\label{eq:phire}
\phi^{\alpha}_{\bold{J}+\bold{1}_i}(x,[u])=D_i\phi_{\bold{J}}^{\alpha}(x,[u])-(D_i\xi^j(x,u))u_{\bold{J}+\bold{1}_j}^{\alpha}.
\end{equation}
 It is often more convenient to equivalently write prolonged generators in terms of the so-called characteristics of symmetries $Q^{\alpha}=\phi^{\alpha}-\xi^iu^{\alpha}_{\bold{1}_i}$, that is
\begin{equation}
\operatorname{pr}\bold{v}=\xi^iD_i+\sum_{\alpha,\bold{J}}(D_{\bold{J}}Q^{\alpha}) \partial_{u_{\bold{J}}^{\alpha}}.
\end{equation}
Here we use the shorthand notation $D_{\bold{J}}=D_1^{j_1}D_2^{j_2}\cdots D_m^{j_m}$ for $\bold{J}=(j_1,j_2,\ldots,j_m)$.
The invariance of a system of local  differential equations 
\begin{equation}
\{F_{k}(x,[u])=0\}_{k=1}^l
\end{equation}
corresponding to the transformations \eqref{eq:lpt} leads to the linearized symmetry condition
\begin{equation}\label{eq:lsc}
\operatorname{pr}\bold{v}(F_{k}(x,[u]))=0,\text{ whenever } \{F_{k}(x,[u])=0\}_{k=1}^l \text{ holds},
\end{equation}
 where $\bold{v}$ is the infinitesimal generator \eqref{eq:inge}.

To extend the above analysis to nonlocal equations of our interest, we define the following reflections for $i=1,2,\ldots,m$, 
\begin{equation}
\operatorname{Ref}^i:(x^1,\ldots,x^i,\ldots,x^m)\mapsto (x^1,\ldots,-x^i,\ldots,x^m),
\end{equation}
and
\begin{equation}
\operatorname{Ref}^i:f(x^1,\ldots,x^i,\ldots,x^m)\mapsto f(x^1,\ldots,-x^i,\ldots,x^m)
\end{equation}
for a function $f$ defined on proper domains. In particular, 
\begin{equation}
\operatorname{Ref}^iu^{\alpha}(x^1,\ldots,x^i,\ldots,x^m)=u^{\alpha}(x^1,\ldots,-x^i,\ldots,x^m),\quad \alpha=1,2,\ldots,n.
\end{equation}
A  system of nonlocal differential equations is then given by 
\begin{equation}
\label{eq:nonlocalde}
\mathcal{A}=
\left\{F_{k}\left(x,[u],[\operatorname{Ref}^iu],[\operatorname{Ref}^i\circ\operatorname{Ref}^ju]_{i< j},\ldots,[u(-x)]\right)=0\right\}_{k=1}^l.
\end{equation}
For simplicity, we will sometimes omit the arguments if they are local variables $x$.
Let us still consider transformations of the form \eqref{eq:lpt} with the infinitesimal generator \eqref{eq:inge}. Now the prolongation formula involving the reflections becomes
\begin{equation}\label{eq:prolong}
\begin{aligned}
\operatorname{pr}_{\operatorname{Ref}}\bold{v}&=\bold{v}+\sum_{\alpha,|\bold{J}|\geq1}\phi_{\bold{J}}^{\alpha}\frac{\partial}{\partial {u_{\bold{J}}^{\alpha}}}\\
&~~~~+\sum_{i,\alpha,|\bold{J}|\geq1}\left(\operatorname{Ref}^i\phi_{\bold{J}}^{\alpha}\right)\frac{\partial}{\partial \left(\operatorname{Ref}^iu_{\bold{J}}^{\alpha}\right)}+\sum_{i< j,\alpha,|\bold{J}|\geq1}\left(\operatorname{Ref}^i\circ \operatorname{Ref}^j\phi_{\bold{J}}^{\alpha}\right)\frac{\partial}{\partial \left(\operatorname{Ref}^i\circ \operatorname{Ref}^ju_{\bold{J}}^{\alpha}\right)}\\
&~~~~+\cdots+\sum_{\alpha,|\bold{J}|\geq1}\phi_{\bold{J}}^{\alpha}\left(-x,[u(-x)]\right)\frac{\partial}{\partial {u_{\bold{J}}^{\alpha}(-x)}},
\end{aligned}
\end{equation}
where the functions $\phi_{\bold{J}}^{\alpha}=\phi_{\bold{J}}^{\alpha}(x,[u])$ are again defined through \eqref{eq:phire}.
Invariance of the nonlocal system \eqref{eq:nonlocalde} with respect to the transformations \eqref{eq:lpt} is equivalent to the linearized symmetry condition that
\begin{equation}\label{eq:lscnonlocal}
\operatorname{pr}_{\operatorname{Ref}}\bold{v} \left(F_{k}\left(x,[u],[\operatorname{Ref}^iu],[\operatorname{Ref}^i\circ\operatorname{Ref}^ju]_{i< j},\ldots,[u(-x)]\right)\right)=0, \text{ whenever  $\mathcal{A}$ holds},
\end{equation}
which is the first order terms about $\varepsilon$ in the Taylor expansions of the nonlocal system \eqref{eq:nonlocalde} evaluated at the new variables $\widetilde{x},\widetilde{u}$ and so forth.

In the next two sections, we will apply this general theory to two integrable nonlocal differential equations, the nonlocal NLS equation and the nonlocal mKdV equation.

\section{The nonlocal NLS equation}
\label{sec:NLS}
An integrable nonlocal NLS equation was proposed  by Ablowitz and Musslimani \cite{AbMu2013}:
\begin{equation}\label{eq:noSc}
\operatorname{i}q_t(x,t)+q_{xx}(x,t)+2q^2(x,t)q^{\ast}(-x,t)=0,
\end{equation}
where $\ast$ denotes complex conjugate and $q(x,t)$ is a complex-valued function of real variables $x$ and $t$.
They showed that it possesses a Lax pair and infinitely many conservation laws, and is solvable via the inverse scattering transform. We study its continuous symmetries in this section.

\subsection{Lie point symmetries}
As $q(x,t)$ is complex-valued, two alternative approaches may be used to calculate its continuous symmetries. Under the coordinate $(x,t,q(x,t),q^{\ast}(x,t))$,  we consider the following local transformations 
\begin{equation}
\begin{aligned}
x&\mapsto x+\varepsilon \xi\left(x,t,q(x,t),q^{\ast}(x,t)\right)+O(\varepsilon^2),\\
t&\mapsto t+\varepsilon \tau\left(x,t,q(x,t),q^{\ast}(x,t)\right)+O(\varepsilon^2),\\
q(x,t)&\mapsto q(x,t)+\varepsilon \phi\left(x,t,q(x,t),q^{\ast}(x,t)\right)+O(\varepsilon^2).
\end{aligned}
\end{equation}
Again, we omit the arguments if they are local variables $(x,t)$. The corresponding infinitesimal generator is 
\begin{equation}
\bold{v}=\xi\left(x,t,q,q^{\ast}\right)\partial_x+\tau\left(x,t,q,q^{\ast}\right)\partial_{t}+\phi\left(x,t,q,q^{\ast}\right)\partial_{q}.
\end{equation}

From Section \ref{sec:theory}, see also \cite{Ku1977,Ol1993}, we know that the prolongation formula for an infinitesimal generator $\xi(x,t,u)\partial_x+\tau(x,t,u)\partial_t+\phi(x,t,u)\partial_u$ of local differential equations is
 \begin{equation}\label{eq:sympro}
\xi D_x+\tau D_t+Q\partial_u+(D_x Q)\partial_{u_x}+(D_t Q)\partial_{u_t}+\cdots+\left(D_x^kD_t^lQ\right)\partial_{\left(D_x^kD_t^lu\right)}+\cdots,
\end{equation}
where the characteristic function is $Q=\phi-\xi u_x-\tau u_t$ and $D^k_x$ denotes $k$ times of total derivatives with respect to $x$ and so forth. For the nonlocal NLS equation, we then adopt the following prolongation formula:
\begin{equation}\label{eq:prco}
\begin{aligned}
\operatorname{pr}_{\operatorname{Ref}}\bold{v}&=\bold{v}+\phi^{\ast}\left(-x,t,q(-x,t),q^{\ast}(-x,t)\right)\partial_{q^{\ast}(-x,t)}+\left(D_t\phi-(D_t\xi)q_x-(D_t\tau)q_t\right)\partial_{q_t}\\
&~~~~+\left(D_x^2\phi-(D_x^2\xi)q_x-2(D_x\xi)q_{xx}-(D_x^2\tau)q_t-2(D_x\tau)q_{tx}\right)\partial_{q_{xx}}+\cdots.
\end{aligned}
\end{equation}
where $D_x$ and $D_t$ denote total derivatives about $x$ and $t$ respectively.
It is generalised from the prolongation formula \eqref{eq:sympro} for symmetries of local differential equations with real variables, but by adding the conjugate terms and their prolongations. 
 
 A vector field $\bold{v}$ generates a group of symmetries for the nonlocal NLS equation if it satisfies  the following   linearized symmetry condition that
 \begin{equation}\label{eq:lscNLS}
 \operatorname{pr}_{\operatorname{Ref}}\bold{v}\left(\operatorname{i}q_t+q_{xx}+2q^2q^{\ast}(-x,t)\right)=0
 \end{equation}
 holds identically for all solutions of the nonlocal NLS equation \eqref{eq:noSc}. We first expanding the left hand side of \eqref{eq:lscNLS} and obtain that
 \begin{equation}
 \begin{aligned}
 \operatorname{i}\left(D_t\phi-(D_t\xi)q_x-(D_t\tau)q_t\right)+D_x^2\phi&-(D_x^2\xi)q_x-2(D_x\xi)q_{xx}-(D_x^2\tau)q_t-2(D_x\tau)q_{tx}\\
 &+4qq^{\ast}(-x,t)\phi+2q^2\phi^{\ast}\left(-x,t,q(-x,t),q^{\ast}(-x,t)\right)=0
 \end{aligned}
 \end{equation}
 restricted to solutions of the nonlocal NLS equation. We then substitute $q_t=\operatorname{i}\left(q_{xx}+2q^2q^{\ast}(-x,t)\right)$ and $q_t^{\ast}=-\operatorname{i}\left(q_{xx}+2q^2q^{\ast}(-x,t)\right)^{\ast}$ inside, leading to a polynomial for $q_x$, $q_x^{\ast}$, $q_{xx}$, $q_{xx}^{\ast}$ and so forth, which equals to zero identically. It is necessary and sufficient for the coefficients of the polynomial to vanish, amounting to a system of partial differential equations for $\xi$, $\tau$ and $\phi$ as follows:
 \begin{equation}
 \begin{aligned}
& D_x\tau=0,~\xi_q=0,~\xi_{q^{\ast}}=0,~\phi_{q^{\ast}}=0,~\phi_{qq}=0,~\tau_t-2\xi_x=0,~\operatorname{i}\xi_t+\xi_{xx}-2\phi_{xq}=0,\\
&\operatorname{i}\phi_t+\phi_{xx}-2\left(\phi_q-\tau_t\right)q^2q^{\ast}(-x,t)+4qq^{\ast}(-x,t)\phi+2q^2\phi^{\ast}\left(-x,t,q(-x,t),q^{\ast}(-x,t)\right)=0.
 \end{aligned}
 \end{equation}
The general solution of the above system is
\begin{equation}
\xi=-C_1x+\operatorname{i}C_2t+C_4,~\tau=-2C_1t+C_5,~\phi=\left(C_1+\operatorname{i}C_3-\frac{1}{2}C_2x\right)q,
\end{equation} 
where $C_1$, $C_2$ and $C_3$ are real-valued while $C_4$ and $C_5$ are complex-valued.
Therefore, the symmetries of the nonlocal NLS equation are generated by the following five infinitesimal generators
\begin{equation}\label{eq:Scsy}
\partial_x,~~\partial_t, ~~\operatorname{i}q\partial_q,~~-x\partial_x-2t\partial_t+q\partial_q, ~~\operatorname{i}t\partial_x-\frac{1}{2}xq\partial_q.
\end{equation} 
They can equivalently be cast into evolutionary type respectively as follows
\begin{equation}
\begin{aligned}
&~~~~~~~~~~-q_x\partial_q,~~\operatorname{i}\left(q_{xx}+2q^2q^{\ast}(-x,t)\right)\partial_q,~~\operatorname{i}q\partial_q,\\
&\left(q+xq_x+2\operatorname{i}t\left(q_{xx}+2q^2q^{\ast}(-x,t)\right)\right)\partial_q,~~\left(-\frac{1}{2}xq-\operatorname{i}tq_x\right)\partial_q.
\end{aligned}
\end{equation}

Alternatively, we can define $q(x,t)=u(x,t)-\operatorname{i}v(x,t)$ where $u(x,t)$ and $v(x,t)$ are real-valued functions, and use the symmetry prolongation formula for real-valued differential equations to calculate symmetries. 
Now the infinitesimal generator is
\begin{equation}
\bold{v}=\xi(x,t,u,v)\partial_x+\tau(x,t,u,v)\partial_t+\phi(x,t,u,v)\partial_u+\eta(x,t,u,v)\partial_v,
\end{equation}
and the equation becomes
\begin{equation}
\left\{
\begin{array}{l}
u_t-v_{xx}-4uvu(-x,t)+2\left(u^2-v^2\right)v(-x,t)=0,\vspace{0.35cm}\\
v_t+u_{xx}+4uvv(-x,t)+2\left(u^2-v^2\right)u(-x,t)=0.
\end{array}
\right.
\end{equation}
 The following symmetries are obtained for the system above, using the linearized symmetry condition \eqref{eq:lscnonlocal} again:
\begin{equation}
\partial_x,~~\partial_t,~~-v\partial_u+u\partial_v, ~~-x\partial_x-2t\partial_t+u\partial_u+v\partial_v.
\end{equation}
They correspond to the first four generators of \eqref{eq:Scsy}. The last one obtained above  does not appear here since it will transform the real-valued $x$ to a complex-valued argument since $\xi=\operatorname{i}t$. 

\subsection{Symmetry reductions}
Next, we will use the symmetries to conduct possible reductions. Preferable, we choose to use the symmetries \eqref{eq:Scsy} with complex variables. The simplest reduction one would expect is probably traveling-wave solutions, which is difficult here as the invariant $x-at$ becomes $-x-at$ at $(-x,t)$. 

Consider the most general infinitesimal generator
\begin{equation}
a\partial_x+b\partial_t+c\operatorname{i}q\partial_q+d\left(-x\partial_x-2t\partial_t+q\partial_q\right)+e\left(\operatorname{i}t\partial_x-\frac{1}{2}xq\partial_q\right)
\end{equation}
where $a,b,c,d,e$ are arbitrary constants. The invariant variables can be found by solving the characteristic equations 
\begin{equation}
\frac{\operatorname{d}\!x}{a-dx+\operatorname{i}et}=\frac{\operatorname{d}\!t}{b-2dt}=\frac{\operatorname{d}\!q}{\operatorname{i}cq+dq-\frac{1}{2}exq},
\end{equation}
and we summarize the results as follows. Note that the equation depends on $q(x,t)$ and $q^{\ast}(-x,t)$ simultaneously, and we must select the constants properly to make the invariants meaningful.
\begin{itemize}
\item If $d=b=0$ (and $a^2 + e^2\neq0$), we have
\begin{equation}
\begin{aligned}
y&=t,\\
q(x,t)&=\exp\left\{\frac{x(4\operatorname{i}c-ex)}{4(\operatorname{i}et+a)}\right\}p(t).
\end{aligned}
\end{equation}
When $a=0$ and $e\neq 0$, the reduced equation is
 \begin{equation}
\operatorname{i}e^2p'(t)+\frac{\operatorname{ie^2}}{2t}p(t)+\frac{c^2}{t^2}p(t)+2|p(t)|^2p(t)=0.
 \end{equation}
\item If $d=0$ and $b\neq 0$, we have
\begin{equation}
\begin{aligned}
y&=bx-\frac{1}{2}\operatorname{i}et^2-at,\\
q(x,t)&=\exp\left\{-\frac{ae}{4b^2}t^2-\frac{e}{2b^2}ty+\operatorname{i}\left(\frac{c}{b}t-\frac{e^2}{12b^2}t^3\right)\right\}p(y).
\end{aligned}
\end{equation}
As $b\neq0$, we must choose $a=e=0$. 
 Next we consider the corresponding reductions to nonlocal and local ordinary differential equations separately here and throughout.
\begin{itemize}
\item {\bf Reduction to a nonlocal ODE.} If we choose $y=x$ and  $q(x,t)=\exp(\operatorname{i}ct)p(y)$, we obtain the nonlocal Painlev\'e-type equation as shown in \cite{AbMu2013}: 
\begin{equation}
p''(y)-cp(y)+2p^2(y)p^{\ast}(-y)=0.
\end{equation}
Note that since $p(y)$ is invariant, so is $p(-y)$; namely the nonlocal invariant is
\begin{equation}
p(-y)=\exp(-\operatorname{i}ct)q(-x,t).
\end{equation}
\item {\bf Reduction to a local ODE.}
Alternatively, we may choose the invariants as $y=x^2$ and $q(x,t)=\exp(\operatorname{i}ct)p(y)$. The reduced equation is a local ODE
\begin{equation}\label{eq:pyy}
4yp''(y)=cp(y)-2p'(y)-2|p(y)|^2p(y).
\end{equation}
If we assume $p(y)$ is real, solution of the above equation can be expressed using the Jacobi elliptic function as
\begin{equation}
p(y)=C_2\sqrt{\frac{c}{C_2^2+c-1}}\operatorname{sn}\left(\sqrt{\frac{c}{C_2^2+c-1}}\left(\sqrt{-(c-1)y}+C_1\right),\frac{C_2}{\sqrt{c-1}}\right),
\end{equation}
where $C_1,C_2$ are integration constants. The above equation can actually be written in a simpler form by introducing $y=z^2$ and $\widehat{p}(z)=p(y)$; the resulting equation is
\begin{equation}
\widehat{p}''(z)=c\widehat{p}(z)-2|\widehat{p}(z)|^2\widehat{p}(z).
\end{equation}
\end{itemize}
\item
 If $d\neq 0$, we have 
\begin{equation}
\begin{aligned}
y&=\frac{d^2x-ad+\operatorname{i}e\left(dt-b\right)}{d^2\sqrt{|2dt-b|}},\\
q(x,t)&=\exp\left\{\left(\frac{ae}{4d^2}-\frac{1}{2}\right)\ln|2dt-b|+\frac{e}{2d}y\sqrt{|2dt-b|}\right\}\times \\
&~~~~~~~~~~~~~~~~\exp\left\{-\operatorname{i}\frac{e^2}{4d^2}t+\operatorname{i}\left(\frac{be^3}{8d^3}-\frac{c}{2d}\right)\ln|2dt-b|\right\}p(y).
\end{aligned}
\end{equation}
Now we must set $a=e=0$. 
\begin{itemize}
\item {\bf Reduction to a nonlocal ODE.} Let 
\begin{equation}
\begin{aligned}
y&=\frac{x}{\sqrt{|2dt-b|}},\\
q(x,t)&=\exp\left\{-\left(\frac{1}{2}+\frac{\operatorname{i}c}{2d}\right)\ln|2dt-b|\right\}p(y).
\end{aligned}
\end{equation}
The reduced equation is a nonlocal ODE
\begin{equation}
p''(y)=\left(\operatorname{i}d-c\right)p(y)+\operatorname{i}dyp'(y)-2p^2(y)p^{\ast}(-y).
\end{equation}
\item {\bf Reduction to a local ODE.}
If we choose the invariant variables via
\begin{equation}
\begin{aligned}
y&=\frac{x^2}{2dt-b},\\
q(x,t)&=\exp\left\{-\left(\frac{1}{2}+\frac{\operatorname{i}c}{2d}\right)\ln|2dt-b|\right\}p(y),
\end{aligned}
\end{equation}
the reduced equation is local, i.e.
\begin{equation}
4yp''(y)=(\operatorname{i}d-c)p(y)+(2\operatorname{i}dy-2)p'(y)-2|p(y)|^2p(y).
\end{equation}
Introducing $y=z^2$ and $\widehat{p}(z)=p(y)$ changes the equation to
\begin{equation}
\widehat{p}''(z)=(\operatorname{i}d-c)\widehat{p}(z)+\operatorname{i}d z\widehat{p}(z)-2|\widehat{p}(z)|^2\widehat{p}(z).
\end{equation}
\end{itemize}
\end{itemize}


%

\section{The nonlocal mKdV equation}
\label{sec:mKdV}
The nonlocal mKdV equation we consider in this paper is, c.f. \cite{JiZh2017},
\begin{equation}
u_t(x,t)+u(x,t)u(-x,-t)u_x(x,t)+u_{xxx}(x,t)=0.
\end{equation}
Assuming that the infinitesimal generator reads
\begin{equation}
\bold{v}=\xi(x,t,u)\partial_x+\tau(x,t,u)\partial_t+\phi(x,t,u)\partial_u,
\end{equation}
its prolongation $\operatorname{pr}_{\operatorname{Ref}}\bold{v}$ can be obtained using \eqref{eq:prolong}.
From a similar procedure for applying the linearized symmetry condition \eqref{eq:lscnonlocal} to the nonlocal NLS equation above, a straightforward calculation gives the following infinitesimal generators for symmetries of the nonlocal mKdV equation:
\begin{equation}
\partial_x, ~~\partial_t,~~ -x\partial_x-3t\partial_t+u\partial_u.
\end{equation}

We follow the same approach as for the nonlocal NLS equation to search for symmetry reductions. The most general symmetry generator can be denoted by
\begin{equation}
a\partial_x+b\partial_t+c\left(-x\partial_x-3t\partial_t+u\partial_u\right), 
\end{equation}
where $a,b,c$ are arbitrary constants. The characteristic equations read
\begin{equation}
\frac{\operatorname{d}\!x}{a-cx}=\frac{\operatorname{d}\!t}{b-3ct}=\frac{\operatorname{d}\!u}{cu}.
\end{equation}
\begin{itemize}
\item When $c=0$, it corresponds to the traveling-wave case.
\begin{itemize}
\item {\bf Reduction to a nonlocal ODE.} The corresponding invariants are
\begin{equation}
\begin{aligned}
y=bx-at \text{ and } v(y)=u(x,t).
\end{aligned}
\end{equation}
The reduced equation is
\begin{equation}\label{eq:waveKdV}
b^3v'''(y)+bv(y)v(-y)v'(y)-av'(y)=0.
\end{equation}
When $b=0$, we obtain constant solution; when $b\neq 0$, without loss of generality, it can be chosen as $b=1$, namely
\begin{equation}
v'''(y)+v(y)v(-y)v'(y)-av'(y)=0.
\end{equation}
 In principle, it can be integrated once as it admits a symmetry generated by $\partial_y$ but it will involve the inverse of nonlocal functions. We will show some of its special solutions with the assumption $a>0$.  
\begin{itemize}
\item Exponential solutions:
\begin{equation}
v(y)=C_1\exp(C_2y) \text{ subject to }  C_1^2+C_2^2=a.
\end{equation}
\item Soliton solutions: 
\begin{equation}
v(y)=\pm \frac{2\sqrt{6a}}{\exp(\sqrt{a}y)+\exp(-\sqrt{a}y)}.
\end{equation}
\end{itemize}
\item {\bf Reduction to a local ODE.} 
We may alternatively introduce the invariants in another way, namely $y=(bx-at)^2$ and $v(y)=u(x,t)$. Now the reduced equation reads
\begin{equation}
4b^3yv'''(y)+6b^3v''(y)+bv^2(y)v'(y)-av'(y)=0,
\end{equation}
which can be integrated once
\begin{equation}
4b^3yv''(y)+2b^3v'(y)+\frac{b}{3}v^3(y)-av(y)+C_1=0.
\end{equation}
This equation can further be simplified by introducing $y=z^2$ and $\widehat{v}(z)=v(y)$, amounting  to
\begin{equation}
b^3\widehat{v}''(z)+\frac{b}{3}\widehat{v}^3(z)-a\widehat{v}(z)+C_1=0.
\end{equation}
The final equation is solvable by letting $\widehat{v}(z)=w(\widehat{v})$; the general solution is
\begin{equation}
z+C_3=\pm \int_0^{\widehat{v}(z)}\frac{\sqrt{6}b^{3/2}}{\sqrt{-bs^4+6as^2-12C_1s+6C_2b^3}}\operatorname{d}\!s,
\end{equation}
where $C_1,C_2,C_3$ are integration constants.
\end{itemize}

\item If $c\neq 0$, the invariants are
\begin{equation}
\begin{aligned}
y=(cx-a)(3ct-b)^{-1/3} \text{ and } v(y)=(3ct-b)^{1/3}u(x,t).
\end{aligned}
\end{equation}
 Now we must set $a=b=0$, namely reduction related to the  generator $-x\partial_x-3t\partial_t+u\partial_u$.  
The related invariants are $y=t^{-1/3}x$ and $v(y)=t^{1/3}u(x,t)$ and we obtain the reduced equation as a local ODE
\begin{equation}
v'''(y)-v^2(y)v'(y)-\frac{v(y)+yv'(y)}{3}=0.
\end{equation}
It can be integrated once to the second Painlev\'e equation 
\begin{equation}
v''(y)=\frac{1}{3}v^3(y)+\frac{1}{3}yv(y)+C.
\end{equation}
\end{itemize}

%
%

We are now able to conclude that all reduced local ODEs for the nonlocal mKdV equation are integrable, that is analogous to the local situation.

\begin{rem}
 In \cite{AbMu2013}, the authors pointed out that similarity reduction of the nonlocal NLS equation may lead to nonlocal ordinary differential equations. However, as shown by the two illustrative examples, such inconvenience can be overcome by choosing the invariant variables/functions in a proper manner and the reduced ODEs become local.
\end{rem}

\section{A Remark on transformations between nonlocal differential equations and differential-difference equations}
\label{sec:ddes}
In \cite{YaYa2017}, authors introduced variable transformations to connect nonlocal and local integrable equations. For instance, the nonlocal NLS equation becomes a local NLS equation under the transformation
\begin{equation}
x= \operatorname{i}\widehat{x}, ~~t= -\widehat{t},~~q(x,t)=\widehat{q}(\widehat{x},\widehat{t}).
\end{equation}
The nonlocal complex mKdV equation  become the local (classical) complex mKdV equation under the transformation
\begin{equation}
x= \operatorname{i}\widehat{x}, ~~t=-\operatorname{i}\widehat{t},~~u(x,t)=\widehat{u}(\widehat{x},\widehat{t}).
\end{equation}
In this section, we will show the relations between nonlocal differential equations and DDEs through variable transformations. 

For the nonlocal NLS equation, we consider the following transformations
\begin{equation}
x= \exp({\widehat{x}}),~~ t=\widehat{t},~~q(x,t)=\widehat{q}(\widehat{x},\widehat{t}),
\end{equation}
where the variable $\widehat{x}$ is imaginary making $x$ imaginary too. Let us drop the  hats (always) and the nonlocal NLS equation becomes a DDE
\begin{equation}
\operatorname{i}q_t+\exp({-2x})\left(q_{xx}-q_x\right)+2q^2q^{\ast}(x+\operatorname{i}\pi,t)=0.
\end{equation}
Let us introduce the following transformations
\begin{equation}
x= \exp({\widehat{x}}),~~ t=\exp({\widehat{t}}),~~u(x,t)=\widehat{u}(\widehat{x},\widehat{t}),
\end{equation}
where the variables $\widehat{x}$ and $\widehat{t}$ are both imaginary. The nonlocal mKdV equation  becomes
\begin{equation}
\begin{aligned}
\exp({-t})u_t&+\exp({-x})uu(x+\operatorname{i}\pi,t+\operatorname{i}\pi)u_x+\exp({-3x})\left(u_{xxx}-3u_{xx}+2u_x\right)=0.
\end{aligned}
\end{equation}
Under the transformation $y=\exp(\widehat{y})$, $v(y)=\widehat{v}(\widehat{y})$, the reduced equation \eqref{eq:waveKdV} becomes  
\begin{equation}\label{eq:wavelocal}
b^3\exp({-2y})\left(v'''(y)-3v''(y)+2v'(y)\right)+\left(bv(y)v(y+\operatorname{i}\pi)-a\right)v'(y)=0.
\end{equation}
These DDEs can further be rescaled and normalized. For example, taking $y=\operatorname{i}\pi \widehat{y}$ and $v(y)=\widehat{v}(\widehat{y})$,
equation \eqref{eq:wavelocal} becomes  
\begin{equation}
b^3\exp(-\operatorname{i}2\pi y)\left(-\frac{1}{\pi^2}v'''(y)+\frac{3\operatorname{i}}{\pi}v''(y)+2v'(y)\right)+\left(bv(y)v(y+1)-a\right)v'(y)=0.
\end{equation}
Similar DDEs were investigated in \cite{QuCaSa1992}, but the variables were real-valued therein. In the same manner, the above DDEs transformed from the nonlocal NLS equation and the nonlocal mKdV equation can also be rescaled, respectively as follows:
\begin{equation}
\operatorname{i}q_t+\exp({-\operatorname{i}2\pi x})\left(-\frac{1}{\pi^2}q_{xx}+\frac{\operatorname{i}}{\pi}q_x\right)+2q^2q^{\ast}(x+1,t)=0.
\end{equation}
and
\begin{equation}
\begin{aligned}
\exp({-\operatorname{i}\pi t})u_t&+\exp({-\operatorname{i}\pi x})uu(x+1,t+1)u_x+\exp({-\operatorname{i}3\pi x})\left(-\frac{1}{\pi^2}u_{xxx}+\frac{3\operatorname{i}}{\pi}u_{xx}+2u_x\right)=0.
\end{aligned}
\end{equation}

\begin{rem}
The above examples showed that simple transformations allow us to transfer nonlocal equations to DDEs. Apparently, similar transformations can be immediately introduced for other nonlocal differential equations/systems using the same manner.
\end{rem}

\section{Conclusions}
In this paper, symmetry analysis was extended to study nonlocal differential equations. The general theory presented in Section \ref{sec:theory} is applicable to any nonlocal differential equations involving space and/or time reflections. In particular, two integrable nonlocal equations, the nonlocal NLS equation and the nonlocal mKdV equation, served as illustrative examples. All Lie point symmetries of these two nonlocal PDEs were obtained and possible symmetry reductions to nonlocal and local ordinary differential equations were conducted. It is shown that, at least for the two illustrative examples, one can always carefully choose the invariant variables to ensure that all reduced differential equations are local.

At the end, we introduced some local transformations transferring nonlocal differential equations to DDEs; it is potential, hence, to extend the existing theories for DDEs to nonlocal differential equations, for instance, symmetries, conservation laws and integrability of DDEs, c.f. \cite{QuCaSa1992,LeWiYa2010,Pe2017,MiWaXe2011,Ya2006,Ku1985}. We will explore more in this direction in a separate project.

\section*{ Acknowledgements} 
This work is partially supported by JSPS Grant-in-Aid for Scientific Research (No. 16KT0024),  the MEXT `Top Global University Project', and Waseda University Grant for Special Research Projects (Nos. 2019C-179, 2019E-036, 2019R-081). 



\begin{thebibliography}{c}
\bibitem{AbMu2017} Ablowitz, M.J.; Musslimani, Z.H.  Integrable nonlocal nonlinear equations. {\it Stud. Appl. Math.}  {\bf 2017}, {\it 139}, 7--59.
\bibitem{AbMu2013} Ablowitz, M.J.; Musslimani, Z.H. Integrable nonlocal nonlinear Schr\"odinger equation. {\it Phys. Rev. Lett.} {\bf 2013}, {\it 110}, 064105, 5pp.
\bibitem{AbMu2016} Ablowitz, M.J.; Musslimani, Z.H. Inverse scattering transform for the integrable nonlocal nonlinear Schr\"odinger equation. {\it Nonlinearity} {\bf 2016}, {\it 29}, 915--946.
\bibitem{AbRaSe1978} Ablowitz, M.J.; Ramani, A.; Segur H. Nonlinear evolution equations and ordinary differential equations of Painlev\'e type. {\it Lett. Nuovo Cimento} {\bf 1978}, {\it 23}, 333--338.
\bibitem{AcHe1975} Ackerman, M.; Hermann, R. {\it Sophus Lie's 1880 Transformation Group Paper}; Math. Sci. Press: Brookline, Mass., USA, 1975.
\bibitem{BlCo1969} Bluman, G.W.; Cole, J.D. General similarity solution of the heat equation. {\it J. Math. Mech.} {\bf 1969}, {\it 18}, 1025--1042.
\bibitem{BlKu1989}  Bluman, G.W.; Kumei, S. {\it Symmetries and Differential Equations}; Springer-Verlag: New York, USA, 1989.
\bibitem{Fo2016}  Fokas, A.S. Integrable multidimensional versions of the nonlocal nonlinear Schr\"odinger equation. {\it Nonlinearity} {\bf 2016}, {\it 29}, 319--324.
\bibitem{GuPe2018} Gurses, M.; Pekcan, A. Nonlocal nonlinear Schr\"{o}dinger equations and their soliton solutions. {\it J. Math. Phys.} {\bf 2018}, {\it 59}, 051501, 17pp.
\bibitem{Hy2000} Hydon, P.E.  {\it Symmetry Methods for Differential Equations: A Beginner's Guide}; Cambridge University Press: Cambridge, UK, 2000.
\bibitem{JiZh2017} Ji, J.L.; Zhu, Z.N. On a nonlocal modified Korteweg-de Vries equation: Integrability, Darboux transformation and soliton solutions. {\it Commun. Nonlinear Sci. Numer. Simulat.} {\bf 2017}, {\it 42}, 699--708.
\bibitem{KhSa2015} Khare, A.; Saxena, A. Periodic and hyperbolic soliton solutions of a number of nonlocal nonlinear equations. {\it J. Math. Phys.} {\bf 2015}, {\it 56}, 032104, 27pp.
\bibitem{Ku1977} Kumei, S. Group theoretic aspects of conservation laws of nonlinear dispersive waves: KdV type equations and nonlinear Schr\"odinger equations. {\it J. Math. Phys.} {\bf 1977}, {\it 18}, 256--264.
\bibitem{Ku1985}  Kupershmidt, B.A. {\it Discrete Lax Equations and Differential-Difference Calculus}; Ast\'erisque, France, 1985.
\bibitem{LeWiYa2010} Levi, D.; Winternitz,  P.; Yamilov, R.I. Lie point symmetries of differential-difference equations. {\it J. Phys. A: Math. Theor.} {\bf 2010}, {\it 43}, 292002, 14pp.
\bibitem{LoHu2017} Lou, S.Y.; Huang, F. Alice-Bob physics: Coherent solutions of nonlocal KdV systems. {\it Sci. Rep.} {\bf 2017}, {\it 7}, 869, 11pp.
\bibitem{MiWaXe2011} Mikhailov, A.V.;  Wang, J.P.; Xenitidis, P. Cosymmetries and Nijenhuis recursion operators for difference equations. {\it Nonlinearity} {\bf 2011}, {\it 24}, 2079--2097.
\bibitem{Ol1993}  Olver, P.J. {\it Applications of Lie Groups to Differential Equations}, 2nd ed.; Springer-Verlag: New York, USA, 1993. 
\bibitem{Pe2017} Peng, L. Symmetries, conservation laws, and Noether's theorem for differential‐difference equations. {\it Stud. Appl. Math.} {\bf 2017}, {\it 139}, 457--502.
\bibitem{QuCaSa1992} Quispel, G.R.W.; Capel, H.W.; Sahadevan, R. Continuous symmetries of differential-difference equations: The Kac--van Moerbeke equation and Painlev\'e reduction. {\it Phys. Lett. A} {\bf 1992}, {\it 170}, 379--383.
\bibitem{SoXiZh2017a} Song, C.Q.;  Xiao, D.M.;  Zhu, Z.N. Reverse space-time nonlocal Sasa--Satsuma equation and its solutions. {\it J. Phys. Soc. Jpn.} {\bf 2017}, {\it 86}, 054001, 6pp.
\bibitem{SoXiZh2017b} Song, C.Q.;  Xiao, D.M.;  Zhu, Z.N.  Solitons and dynamics for a general integrable nonlocal coupled nonlinear Schr\"odinger equation. {\it Commun. Nonlinear Sci. Numer. Simul.} {\bf 2017}, {\it 45}, 13--28.
\bibitem{XuCh2016} Xu, Z.X.; Chow, K.W. Breathers and rogue waves for a third order nonlocal partial differential equation by a bilinear transformation. {\it Appl. Math. Lett.} {\bf 2016}, {\it 56}, 72--77.
\bibitem{Ya2006} Yamilov, R. Symmetries as integrability criteria for differential difference equations. {\it J. Phys. A: Math. Gen.} {\bf 2006}, {\it 39}, R541--R623.
\bibitem{Ya2015} Yan, Z. Integrable $\mathcal{PT}$-symmetric local and nonlocal vector nonlinear Schr\"odinger equations: A unified two-parameter model. {\it Appl. Math. Lett.} {\bf 2015}, {\it 47}, 61--68.
\bibitem{YaYa2017} Yang, B.; Yang, J. Transformations between nonlocal and local integrable equations. {\it Stud. Appl. Math.} {\bf 2017}, {\it 140}, 178--201.
 \bibitem{Zh2018} Zhou, Z.X.  Darboux transformations and global solutions for a nonlocal derivative nonlinear Schr\"odinger equation. {\it Commun. Nonlinear Sci. Numer. Simul.} {\bf 2018}, {\it 62}, 480--488.
\end{thebibliography}
\end{document}